\documentclass[aps,prb,twocolumn,superscriptaddress,showpacs]{revtex4}
\usepackage{amsmath}
\usepackage{bm}
\usepackage{graphicx}

\begin{document}

\title{Controlling Fano and Dicke effects via a magnetic flux in a
double quantum-dot molecule}
\author{P.\ A.\ Orellana}
\affiliation{Departamento de F\'{\i }sica, Universidad
Cat\'{o}lica del Norte, Casilla 1280, Antofagasta, Chile}
\author{M. L. Ladr\'on de Guevara}
\affiliation{Departamento de F\'{\i }sica, Universidad
Cat\'{o}lica del Norte, Casilla 1280, Antofagasta, Chile}
\author{ F. Claro}
\affiliation{Departamento de F\'{\i }sica, Pontificia Universidad
de Chile, Casilla 306 Santiago 22, Chile}

\begin{abstract}
The electronic transport through a parallel double quantum-dot
molecule attached asymmetrically to leads is studied under a
magnetic field. We model the system by means of a non interacting
two-impurity Anderson Hamiltonian. We find that the conductance
shows Fano and Dicke effects that can be controlled by the
magnetic flux.
\end{abstract}


\maketitle

\section{Introduction}

Resonant tunneling through two parallel quantum dots has attracted much
interest recently. For instance, Holleitner \emph{et al.} \cite{holleitner1}
studied how the molecular states of semiconductor quantum dots connected in
parallel to the leads can be coherently probed and manipulated in transport
experiments, while Kubala \emph{et al.}\cite{kubala} reported a level
attraction in an Aharonov-Bohm interferometer with two quantum dots in its
arms. Moreover, Kang \emph{\ et al.}\cite{kang} and Boese \emph{et al.}\cite
{boese} studied the double quantum dot molecule in the parallel geometry in
the presence of a magnetic flux.

In Ref. \onlinecite{lldg} we reported on the transition from a series to a
parallel arrangement of a quantum dot molecule attached to leads. The existence of two
different pathways for the electron transport produces conductance spectra
composed by a Breit-Wigner resonance and a Fano-like resonance at the
bonding and antibonding frequencies, respectively. A progressive line
narrowing (widening) of the Fano (Breit-Wigner) peak is observed as the
system transits from the configuration in series to a symmetrical parallel
one. When the symmetry is total, the Fano line shape is suppressed,
indicating the cancelation of tunneling through the antibonding state, while
the other peak doubles its width. The general features of the conductance
spectrum taking place in the series to parallel transition of Ref. %
\onlinecite{lldg} are given in a parallel molecule embedded in an
Aharonov-Bohm flux, as discussed by Kang \emph{et al.} \cite{kang} and Bai
\emph{et al.}. \cite{bai} The two conductance peaks (Breit-Wigner and
Fano-like curves) depend sensitively on the external magnetic field, and
exhibit Aharonov-Bohm-type oscillations.

In this we consider electron transport through a parallel
quantum-dot molecule embedded in an Aharonov-Bohm interferometer connected
asymmetrically to leads. We show that with a period of a quantum of flux ($\Phi_0=h/e$)the magnetic filed allows interchanging
the roles of
the bonding and antibonding states in the transmission spectrum. For
intermediate values of the flux (namely, semi-integer multiples of a quantum of
flux) the parallel molecule behaves as if it were connected in series. We also find that whenever the flux is close to integer
multiples of $\Phi_0$, the density of states shows an ultranarrow and a broad
peak at the energies of the molecular states, associated to Fano and
Breit-Wigner lineshapes in the conductance. When the flux has exactly the
above values, the conductance experiences the suppression of the Fano line
shape, indicating a localization of the corresponding molecular state,
similarly to what takes place for the symmetrical case in the absence of
magnetic field.\cite{lldg} We find that these results hold even under a
strong left-right asymmetry. This phenomenon resembles the Dicke effect in
optics, which takes place in the spontaneous emission of a pair of atoms
radiating a photon with a wave length much larger than the separation
between them. \cite{dicke} The luminescence spectrum is characterized by a
narrow and a broad peak, associated with long and short-lived states,
respectively. The former state, coupled weakly to the electromagnetic field,
is called \emph{subradiant}, and the latter, strongly coupled, \emph{%
superradiant} state.

The appearance of the Dicke effect in resonant tunneling was first obtained
by Shahbazyan and Raikh in a work on a tunneling junction with two
impurities.\cite{shahbazyan} Later, Shahbazyan and Ulloa \cite{ulloa}
studied this effect in a system of localized states in a strong magnetic
field. More recently, Vorrath and Brandes \cite{vorrath} studied the
stationary current through a double quantum dot interacting via a common
phonon environment.


\section{Model}

\begin{figure}[h]
\centering
\includegraphics[scale=0.3,angle=0]{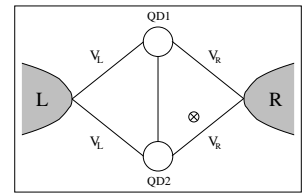}
\caption[Fig1]{Double quantum-dot molecule connected in parallel
to leads} \label{Fig1}
\end{figure}

We consider two single-level quantum dots forming a molecule, coupled to
left and right leads in the way shown in Fig. \ref{Fig1}. The interdot and
intradot electron-electron interactions are neglected.

\noindent The system can be modeled by a non-interacting two-impurity
Anderson Hamiltonian
\begin{equation}
H=H_{M}+H_{L}+H_{I}.  \label{eq-1}
\end{equation}
\noindent Here $H_{M}$ describes the dynamics of the isolate molecule, and
is given by
\begin{equation}
H_{M}=\sum_{i=1}^{2}\varepsilon _{i}d_{i}^{\dag }d_{i}-t_{c}(d_{2}^{\dag
}d_{1}+d_{1}^{\dag }d_{2}),  \label{eq-2}
\end{equation}
where $\varepsilon _{i}$ is the energy of dot $i$, $d_{i}$ $(d_{i}^{\dagger
})$ annihilates (creates) an electron in dot $i$ and $t_{c}$ is the interdot
tunneling coupling parameter. $H_{L}$ is the Hamiltonian for the
noninteracting electrons in the left and right leads
\begin{equation}
H_{L}=\sum_{k_{\alpha }\in \{L,R\}}\varepsilon _{k_{\alpha }}c_{k_{\alpha
}}^{\dag }c_{k_{\alpha }},  \label{eq-3}
\end{equation}
where $c_{k_{\alpha }}$ $(c_{k_{\alpha }}^{\dag })$ is the annihilation
(creation) operator of an electron of momentum $k_{\alpha }$ and
energy $\varepsilon _{k_{\alpha }}$ in the contact $\alpha $. Finally, $%
H_{I} $ accounts for the tunneling between dots and leads,
\begin{eqnarray}
H_{I} &=&\sum_{k_{\alpha }\in \{L,R\}}V_{1k_{\alpha }}d_{1}^{\dag
}c_{k_{\alpha }}+\mbox{ h. c.}  \nonumber \\
&&+\sum_{k_{\alpha }\in \{L,R\}}V_{2k_{\alpha }}d_{2}^{\dag }c_{k_{\alpha }}+%
\mbox{ h. c.},  \label{eq-4}
\end{eqnarray}
with $V_{ik_{\alpha }}$ the tunneling matrix element.

We focus in the density of states and conductance at zero temperature, which
are calculated via the equation of motion approach for the Green's function
\cite{kubala}. The retarded Green's function is defined by
\begin{equation}
G_{ij}^{r}(t)=-i\theta (t)\langle \{d_{i}(t),d_{j}^{\dagger }(0)\}\rangle
,\quad i,j=1,2,  \label{eq-6}
\end{equation}
where $\theta(t)$ is the step function. The linear conductance $G$ is
related to the transmission $T(\varepsilon )$ of an electron of energy $%
\varepsilon $ by the Landauer formula at zero temperature \cite{kubala},
\begin{equation}
G=\frac{2e^{2}}{h}T(\varepsilon _{F}).  \label{eq-5}
\end{equation}
In the absence of interaction, the total transmission $T(\varepsilon )$ can
be expressed as
\begin{equation}
T(\varepsilon )=tr\{\mathbf{G}^{a}(\varepsilon )\mathbf{\Gamma }^{R}\mathbf{G%
}^{r}(\varepsilon )\mathbf{\Gamma }^{L}\},  \label{eq-7}
\end{equation}
where $\mathbf{G}^{r(a)}(\varepsilon )$ is the Fourier transform of the
retarded (advanced) Green's function of the molecule, and the matrix $%
\mathbf{\Gamma }^{L(R)}$ describes the tunneling coupling of the two quantum
dots to the left (right) lead, with
\begin{equation}
\Gamma _{ij}^{L(R)}=2\pi \sum_{k_{L(R)}}V_{ik_{L(R)}}V_{jk_{L(R)}}^{*}\delta
(\varepsilon -\varepsilon _{k_{L(R)}}),\quad i,j=1,2.  \label{eq-8}
\end{equation}

We consider the case of quantum dots with equal energy levels: $%
\varepsilon_1=\varepsilon_2\equiv\varepsilon_0$, and denote by $|\pm\rangle$
the eigenstates of the Hamiltonian of the isolated molecule [Eq. \ref{eq-2}%
], $|\pm\rangle=(|1\rangle\mp|2\rangle)/2$, with energies $%
\varepsilon_\pm=\varepsilon_0+\pm t_c$. We assume that the
magnitudes of the tunneling matrix elements between the dots and
the left and right leads are such that
$|V_{1k_L}|=|V_{2k_L}|=V_L$, and $|V_{1k_{R}}|=|V_{2k_R}|=V_R$,
so that in the presence of a magnetic flux, they have the form $%
V_{1k_L}=V_Le^{-i\phi/4}$, $V_{2k_L}=V_L e^{i\phi/4}$ $V_{1k_R}=V_R
e^{i\phi/4}$,$V_{2k_R}=V_R e^{-i\phi/4}$. Thus, in the basis $%
\{|-\rangle,|+\rangle\}$,
\begin{equation}
{\mathbf \Gamma}^{L,R}=\Gamma^{L,R}\left(
\begin{array}{ll}
\;\;\;\; 1 & e^{\mp i \phi/2} \\
e^{\pm i \phi/2} & \;\;\;\; 1
\end{array}
\right),
\end{equation}
where we have used Eq. (\ref{eq-8}).

For simplicity, we set $\varepsilon_0=0$. The retarded Green's function in
the same basis is
\begin{equation}
\mathbf{G}^{r}(\varepsilon)=\frac{1}{\Omega }\left(
\begin{array}{ll}
\varepsilon -t_c+i\Gamma _{++}/2 & \;\;\;\;-i \Gamma _{-+}/2 \\
\;\;\;\;-i\Gamma _{+-}/2 & \varepsilon +t_c+i\Gamma _{--}/2
\end{array}
\right),  \label{eq-9}
\end{equation}
where
\begin{equation}
\Omega =(\varepsilon +t_c+i\frac{\Gamma_{--}}{2})(\varepsilon -t_c+i\frac{%
\Gamma _{++}}{2})+\frac{\Gamma _{-+}\Gamma _{+-}}{4},  \label{eq-10}
\end{equation}
with $\Gamma_{--}=2 \tilde{\Gamma} \cos^2(\phi/4)$, $\Gamma_{++}=2\tilde{%
\Gamma}\sin^2(\phi/4)$, $\Gamma_{-+}=i(\Gamma^L-\Gamma^R)\sin(\phi/2)$, $%
\Gamma_{+-}=(\Gamma_{-+})^*$, where $\tilde{\Gamma}=\Gamma^L+\Gamma^R$. From
the diagonal elements of the Green's function we can get the spectral
densities $A_{\pm}= -(1/\pi)\mbox{ Im }G_{\pm\pm}^r$. Summing over the $\pm$
states we obtain the density of states of the quantum-dot molecule,
\begin{equation}
\rho(\varepsilon)=\sum\limits_{\sigma =-,+}A_{\sigma},
\end{equation}
where
\begin{equation}
A_-=\frac{1}{\pi\Lambda}\cos^2(\phi/4)\tilde{\Gamma}[(t_c-\varepsilon)^2+
4\Gamma^L\Gamma^R \sin^4(\phi/4)],
\end{equation}
\begin{equation}
A_+=\frac{1}{\pi\Lambda}\sin^2(\phi/4)\tilde{\Gamma}[(\varepsilon+t_c)^2+
4\Gamma^L \Gamma^R \cos^4(\phi/4)]
\end{equation}
with
\begin{equation}
\Lambda=\tilde{\Gamma}^2[\varepsilon-t_c\cos(\phi/2)]^2
+[(t_c+\varepsilon)(t_c-\varepsilon)+\Gamma^L\Gamma^R\sin^2(\phi/2)]^2.
\end{equation}
The transmission, in turn, is given by
\begin{equation}
T(\varepsilon)=\frac{1}{\Lambda}4\Gamma^L\Gamma^R[t_c-\varepsilon\cos(%
\phi/2)]^2.
\end{equation}
When $\phi=0$ and $\Gamma^L=\Gamma^R=\Gamma_0$ (with $\Gamma_0$ the level
broadening of a single quantum dot) we recover the result without magnetic
field. \cite{lldg} The density of states is the sum of a Lorentzian
with width $\Gamma_{-}\rightarrow \tilde{\Gamma}=\Gamma_L+\Gamma_R=2\Gamma_0$
and a delta function centered at the antibonding energy ($%
\Gamma_{+}\rightarrow 0$). In other words, in this limit the antibonding
state is decoupled from the continuum, while the bonding state has reduced
its lifetime to a half. This is due to quantum interference in the
transmission through the two different discrete states (the two quantum-dot
levels) coupled to common leads. This result is totally analogous to the
Dicke effect in optics, that takes place in the spontaneous emission of two
closely-lying atoms radiating a photon into the same environment.\cite{dicke}

The Dicke effect in the parallel molecule occurs not only for the
symmetrical case, but also when $\Gamma^L\neq\Gamma^R$, and whenever the
magnetic flux approaches to an integer of flux quanta. In fact, when $%
\phi=2\pi n$, with $n$ integer, the density of states takes the form
\begin{eqnarray}
\rho (\varepsilon)&=&\frac{1}{\pi}\left[ \frac{\tilde{\Gamma} }{(\varepsilon
-\varepsilon _{-})^{2}+\tilde{\Gamma}^{2}}\right]+
\delta(\varepsilon-\varepsilon_{+}), \quad \mbox{n
even}  \nonumber \\
 &=&\delta(\varepsilon-\varepsilon_{-})+\frac{1}{\pi }\left[ \frac{\tilde{%
\Gamma}}{(\varepsilon -\varepsilon _{+})^{2}+\tilde{\Gamma}^{2}}
\right],\quad \mbox{n odd}.
\end{eqnarray}
In other words, the positions of the long-lived and the short-lived states
can be interchanged depending on the parity of the magnetic flux. Notice that when $t_c=0$ the Dicke effect is still valid for $\phi=2\pi n$,
but when the system is degenerate ($\varepsilon_{+}=\varepsilon_{-}=0$) the
narrow and the wide peaks in the density of states are superimposed, as we
show below.

In order to evaluate the above expressions numerically, we set $%
\Gamma_L=(1-\Delta_A)/\Gamma_0$,$\Gamma_R=(1+\Delta_A)/\Gamma_0$, where $%
\Delta_A$ is the parameter of asymmetry. Figure 2 shows the
density of states for the cases discussed above for
$\Delta_A=0.5$. Fig 2 (a) shows that density of states for
$t_c=\Gamma_0$ (solid line) and $t_c=0$ (dash line) at
$\phi=0.1\pi$. For $t_c=\Gamma_0$, we see that the narrow peak
(corresponding to the "subradiant" state) developes around the
antibonding state, and the broad peak (corresponding to the
"superradiant" state) developes around the bonding state. In the
case with $t_c=0$, both the broad and narrow peaks are centered at
$\varepsilon=0$. For $\phi=1.9 \pi$ and $t_c=\Gamma_0$ the two
peaks interchange roles as displayed in Fig 2 (b), while at
$t_c=0$ the two peak are superimposed at $\varepsilon=0$.
\begin{figure}[h]
\centering
\includegraphics[scale=0.3,angle=0]{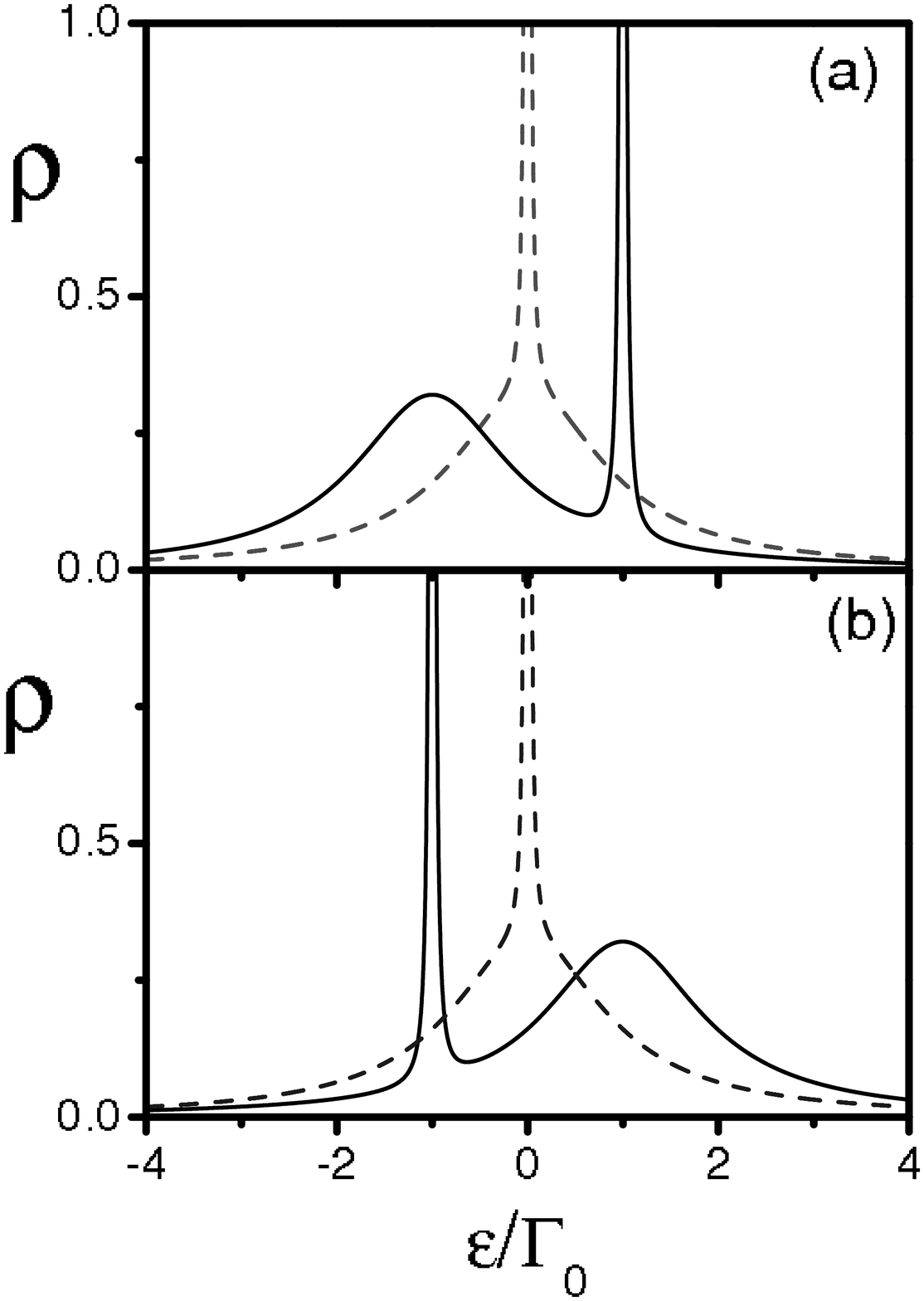}
\caption[Fig2]{Density of states $\rho $ as a function of the
Fermi energy} for a) $t_c=\Gamma_0$ (solid line) and $t_c=0$ (dash
line) at $\phi=0.1$ and b) $t_c=\Gamma_0$ (solid line) and $t_c=0$
(dash line) at $\phi=1.9\pi$ \label{Fig2}
\end{figure}

Next we show that the Dicke effect is also present in the conductance. In
dimensionless units conductance ($g=G/(2e^{2}/h)$) this quantity is given by
\begin{equation}
g(\varepsilon )=\frac{4\Gamma ^{L}\Gamma ^{R}}{\Lambda }[\varepsilon \cos
(\phi /2)-t_{c}]^{2}.
\end{equation}
\noindent In general the conductance spectrum is composed of
Breit-Wigner and Fano line shapes, as shown previously by Kang et
al \cite {kang} and us. \cite{lldg} We note that the width of the
Breit-Wigner and Fano line shapes can be controlled with the
magnetic flux. In fact when the magnetic flux $\phi \rightarrow
2\pi n$, for $n$ even (odd) the Fano line shape is associated with
the antibonding (bonding) state. In the limit $\phi =2\pi n$, the
Fano line shape is suppressed and only the Breit-Wigner line shape
survives. It develops around the bonding or the antibonding
energies depending on whether $n$ is even or odd, respectively,

\begin{equation}
g_{\pm }=\frac{4\Gamma _{L}\Gamma _{R}}{\left[ (\varepsilon -\varepsilon
_{\pm })^{2}+\tilde{\Gamma}^{2}\right] }
\end{equation}
\noindent

The above cases are displayed in Fig. 3 for $\Delta _{A}=0.5$. The curves $g$
versus Fermi energy for $\phi =0.1\pi $ (solid line) and $\phi =0$ (dash
line) are shown in Fig 3 (a). Figure 3 (b) gives the $g$ versus Fermi energy
for the magnetic flux $\phi =1.9\pi $ (solid line) and $\phi =2\pi $ (dashed
line). The above result can be interpreted in the following way. When the
magnetic flux is an integer number flux quanta, the long-lived state is
decoupled from the continuum and is supressed from transmission. For a flux
close to any of these points the system would be in a regime of Dicke effect.

\begin{figure}[h]
\centering
\includegraphics[scale=0.3,angle=0]{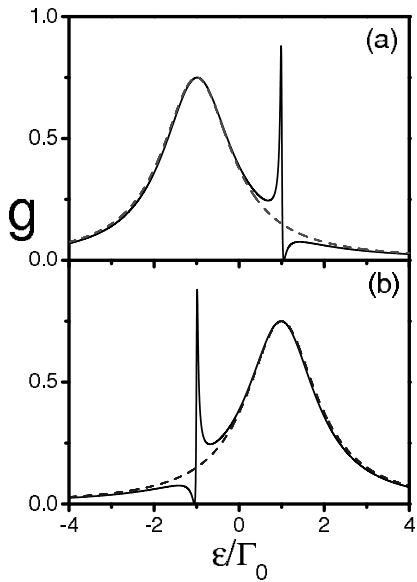}
\caption[Fig3]{Dimensionless conductance $g$ as a function of the Fermi
energy for $t_{c}=\Gamma _{0}$. (a) $\phi =0.1\pi $ (solid line) and $\phi =0
$ (dash line), and (b) $\phi =1.9\pi $ (solid line) and $\phi =2\pi $ (dash
line).}
\label{Fig3}
\end{figure}

We consider now the special case when the quantum dots are disconnected from
each other ($t_{c}=0$). In this case the Fano antiresonance is localized at $%
\varepsilon =0$ independent of the magnetic flux, except for $\phi =2\pi n$.
Therefore, for $\varepsilon =0$ the conductance takes the values

\begin{eqnarray}
g &=&0,\quad \quad \quad \quad \quad (\phi \neq 2\pi n,\mbox{
 $n$ integer}) \\
g &=&\frac{4\Gamma _{L}\Gamma _{R}}{(\Gamma _{L}+\Gamma _{R})^{2}}=1-\Delta
_{A}^{2},\quad (\phi =2\pi n,\mbox{ $n$ integer}).  \nonumber
\end{eqnarray}

\noindent The conductance is different from zero periodically in the
magnetic field, with a period of one quantum of flux. Note also that when $%
\phi =2\pi (n+1/2)$ with $n$ integer, the Lorentzians have the same widths ($%
\Gamma _{+}=\Gamma _{+}=\sqrt{2}\Gamma $). For $\phi \rightarrow 2\pi (n+1/2)
$ ($n$ integer), $cos(\phi )\rightarrow 0$ and hence the behavior of the
conductance is reduced to a the convolution of two Breit-Wigner line shapes
with the same width centered in the bonding and antibonding energies,
respectively,
\begin{widetext}
\begin{equation}
g=\frac{\Gamma_L\Gamma_R t_c^{2}}{\left[ (\varepsilon
-i\Gamma_L)(\varepsilon -i\Gamma_R)-t_c^2\right] \left[
(\varepsilon +i\Gamma_L)(\varepsilon +i\Gamma_R)-t_c^2\right] }.
\end{equation}
\end{widetext}
No Fano line shape develops. The double-quantum dot in the parallel
configuration behaves as a serial one for the transmission. When the
electron crosses the upper (lower) arm, it accumulates a phase difference $%
\pi /2$ (-$\pi /2$). The contribution to the wave function of both paths
interfere destructively and cancel mutually at the leads. Therefore, the
paths that contribute to the conductance are only those that cross the
molecule through both quantum dots sequentially, as in a serial
configuration. Note that when $t_{c}=0$ the conductance vanishes
independently of the energy and perfect reflection is reached. A similar
result was obtained previously by Kubala and K\"{o}nig for a parallel double
quantum dot system connected symmetrically to the leads.\cite{kubala}

\begin{figure}[t]
\centering
\includegraphics[scale=0.3,angle=0]{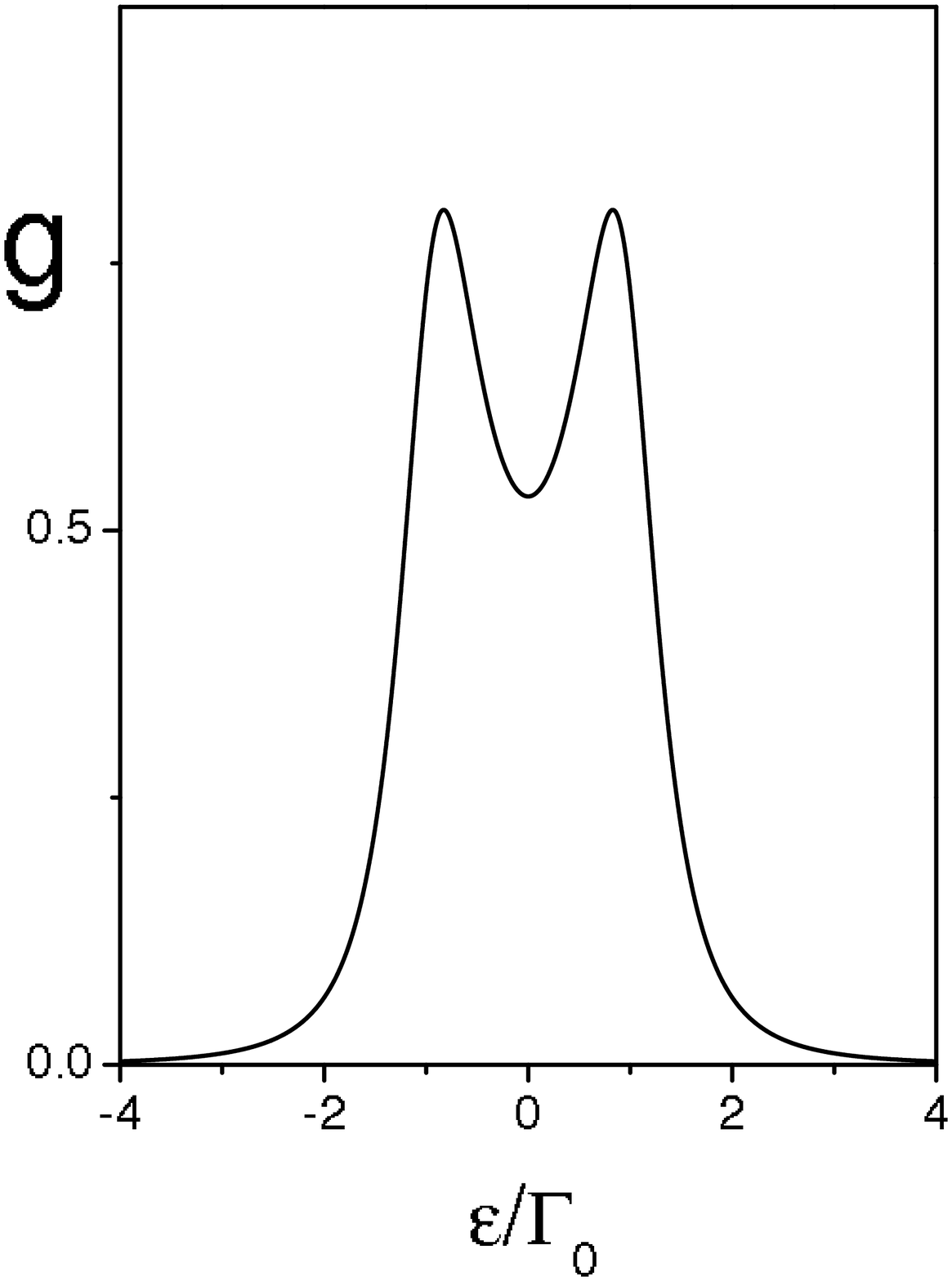}
\caption[Fig4]{Dimensionless conductance $g $ as a function of the
Fermi energy for $t_c=\Gamma_0$ and $\phi=\pi$.}  \label{Fig4}
\end{figure}

\section{Conclusions}

In this work, we studied the conductance and the density of states
at zero temperature of a quantum dot molecule connected
asymmetrically to leads in a parallel configuration under a
magnetic flux. We show that the magnetic flux can control the
different regimes of conduction through the system. In particular,
when the magnetic flux is close an integer number of  flux quanta,
the system is in the Dicke regimen. The conductance spectrum is
composed of Breit-Wigner and Fano line shapes at the bonding and
antibonding energies, or vice versa, depending on whether this
number is even or odd, with their line broadenings controlled by
the magnetic flux. The narrowing (broadening) of a line in the
conductance can be interpreted as an increase (reduction) of the
lifetime of the corresponding molecular state. From the densities
of states it can be deduced that the antibonding (bonding) state
becomes progressively localized as the magnetic flux tends to an
integer number of flux quanta. When the magnetic flux is exactly a
integer number of quantum flux the tunneling through the
antibonding (bonding) state is totally suppressed and the bonding
(antibonding) is the only participating in the transmission.
Moreover, when the magnetic flux is a half integer of flux quanta,
the double-quantum dot in the parallel configuration behaves as a
serial one for the conductance. The control of the decoherence
processes with the magnetic field exhibited by the present system
may have applications in quantum computing.

\begin{acknowledgments}
The authors would like to thank financial support from FONDECYT
under grants 1020269, 1020829 and 1040385. P.A.O. also thanks to
Milenio ICM P02-054-F and M. L. L. d. G. receives financial
support from Universidad Cat\'olica del Norte (internal grant
DGIP).
\end{acknowledgments}


\begin{thebibliography}{99}


\bibitem{holleitner1} A.\ W.\ Holleitner, C.\ R.\ Decker, H.\ Qin, K.\ Ebert, and
R.\ H.\ Blick, Phys.\ Rev.\ Lett.\ \textbf{87}, 256802 (2001).

\bibitem{kubala}  Bj\"{o}rn Kubala and J\"{u}rgen K\"{o}nig, Phys. Rev. B
\textbf{65}, 245301.

\bibitem{kang}  Kicheon Kang and Sam Young Cho, Cond-mat/0210009v1 (unpublished).

\bibitem{boese} Daniel Boese, Walter Hofstetter, Herbet Schoeller, Phys. Rev. B
\textbf{66} 125315 (2002).



\bibitem{lldg} M. L. Ladr\'on de Guevara, F. Claro, and Pedro A.
Orellana, Phys. Rev. B \textbf{67}, 195335 (2003).

\bibitem{bai} Zhi-Ming Bai, Min-Fong Yang, and Yung-Chung Chen,
cond-mat/0307071.

\bibitem{dicke} R. H. Dicke, Phys. Rev. {\textbf 93}, 99 (1954).

\bibitem{shahbazyan} T. V. Shahbayzan, and M. E. Raikh, Phys. Rev.
B {\textbf 49}, 17123 (1994).

\bibitem{ulloa} T. V. Shahbayzan, and S.E. Ulloa, Phys. Rev.
B {\textbf 57}, 6642 (1998).

\bibitem{vorrath} T. Vorrath, and T. Brandes, Phys. Rev.
B {\textbf 68}, 035309 (2003).




\end{thebibliography}
\end{document}